\documentclass[twoside,12pt]{article}
\usepackage{graphicx}
\usepackage{amssymb,amsmath}
\usepackage{bm}
\usepackage{color}

\def\Journal#1#2#3#4{{#1} {#2} (#4) #3}
\def\AA{{\em Astron. Astrophys.}}
\def\APJ{{\em Astrophys. J.}}
\def\APJS{{\em Astrophys. J. Suppl.}}
\def\MNRAS{{\em Mon. Not. Roy. Astron. Soc.}}
\def\ARNPS{{\em Ann. Rev. Nucl. Part. Sci.}}
\def\NCA{{\em Nuovo Cimento} A}

\def\Nature{{\em Nature}}
\def\NPA{{\em Nucl. Phys.} A}
\def\NPB{{\em Nucl. Phys.} B}

\def\NPB{{\em Nucl. Phys.} B}

\def\PLB{{\em Phys. Lett.} B}

\def\PRL{\em Phys. Rev. Lett.}
\def\PREV{\em Phys. Rev.}

\def\PRD{{\em Phys. Rev.} D}
\def\PRC{{\em Phys. Rev.} C}
\def\PRB{{\em Phys. Rev.} B}

\def\IJMPA{{\em Int. J. Mod. Phys.} A}
\def\JHEP{{\em J. High Energy Phys. }}
\def\JPG{{\em J. Phys.} G}
\def\LTP{{\em Low Temp. Phys. }}

\newcommand{\be}{\begin{equation}}
\newcommand{\ee}{\end{equation}}
\newcommand{\bea}{\begin{eqnarray}}
\newcommand{\eea}{\end{eqnarray}}

\begin{document}
%%Preprint: UWO-TH-11/14
\title{ \vspace{1cm} Surprises in relativistic matter in a magnetic field}
\author{E. V. Gorbar,$^{1,2}$ V. A. Miransky,$^{3}$ I. A. Shovkovy,$^{4}$\\
\\
$^1$Department of Physics, Taras Shevchenko National Kiev University,\\ 03022, Kiev, Ukraine\\
$^2$Bogolyubov Institute for Theoretical Physics,\\ 03680, Kiev, Ukraine\\
$^3$Department of Applied Mathematics, University of Western Ontario, \\ London, Ontario N6A 5B7, Canada\\
$^4$Department of Applied Sciences and Mathematics, \\ Arizona State University, Mesa, Arizona 85212, USA}
\maketitle
\begin{abstract} 
A short review of recent advances in understanding the dynamics of relativistic matter in a magnetic field
is presented. The emphasis is on the dynamics related to the generation of the chiral shift parameter in 
the normal ground state. We argue that the chiral shift parameter contributes to the axial current density, 
but does not modify the conventional axial anomaly relation. The analysis based on gauge invariant 
regularization schemes in the Nambu-Jona-Lasinio model suggests that these findings should be 
valid also in gauge theories. It is pointed out that the chiral shift parameter can affect observable 
properties of compact stars and modify the key features of the chiral magnetic effect in heavy ion 
collisions.
\end{abstract}
%\eject
%\tableofcontents
\section{Introduction}

Many recent theoretical studies \cite{Son-et-al,Metlitski:2005pr,Kharzeev:2007tn,Gorbar:2009bm,
Rebhan,FI1,Basar:2010zd,Kim,Frolov,Osipov:2007je,Gatto:2010qs,Mizner,Fayazbakhsh:2010bh} 
of relativistic matter under extreme conditions, e.g., those realized inside compact stars and/or in 
heavy ion collisions, revealed that matter in a strong magnetic field may hold some new surprises.
(For lattice studies of relativistic models in strong magnetic fields, see 
Refs.~\cite{Buividovich:2009wi,D'Elia,arXiv:1111.4956}.) Here we discuss one of them, which 
was triggered by a finding that a topological contribution to the axial current is already induced at 
the lowest Landau level (LLL) in the free theory in a magnetic field \cite{Metlitski:2005pr}. This led 
us to propose that in the realistic case, with interactions, the ground state of such a matter is 
characterized by a chiral shift parameter $\Delta$ \cite{Gorbar:2009bm}, which enters the effective 
Lagrangian density through the $\Delta \bar\psi \gamma^3 \gamma^5 \psi$ term and is generated 
at all Landau levels. The value of $\Delta$ determines a relative shift of the longitudinal momenta 
in the dispersion relations of opposite chirality fermions, $k^{3}\to k^{3}\pm\Delta$, where $k^{3}$ 
is the momentum along the magnetic field. This conclusion is approximately valid even in the case 
of massive particles, if the (ultra-)relativistic regime is realized. This is achieved, for example, in 
matter at a sufficiently high density (i.e., $\mu\gg m$, where $\mu$ is the chemical potential and 
$m$ is the mass of fermions), or a sufficiently high temperature ($T\gg m$) \cite{Gorbar:2011ya}.

The chiral shift parameter is even under the parity transformation ${\cal P}$ and the charge conjugation 
${\cal C}$, but breaks the time reversal ${\cal T}$ and the rotational symmetry $SO(3)$ down to $SO(2)$. 
Since the global symmetries of dense relativistic matter in an external magnetic field are exactly the 
same, the generation of the chiral shift parameter is expected even in perturbation theory \cite{Gorbar:2009bm}. 

\section{Main results}

{\bf Model.} 
Here we briefly review the dynamics responsible for the generation of the chiral shift parameter. 
While we use a simple Nambu-Jona-Lasinio model (NJL), we also envision the generalization 
of the main results to gauge theories. Keeping this in mind, we will utilize a gauge-invariant 
regularization scheme in the analysis below. 

The Lagrangian density of the model reads
\begin{equation}
{\cal L} = \bar\psi \left(iD_\nu+\mu_0\delta_{\nu}^{0}\right)\gamma^\nu \psi
-m_{0}\bar\psi \psi +\frac{G_{\rm int}}{2}\left[\left(\bar\psi \psi\right)^2
+\left(\bar\psi i\gamma^5\psi\right)^2\right],
\label{NJLmodel}
\end{equation}
where $m_{0}$ is the bare fermion mass, $\mu_0$ is the chemical potential, and $G_{\rm int}$ is a 
dimensional coupling constant. By definition, $\gamma^5\equiv i\gamma^0\gamma^1\gamma^2\gamma^3$. 
The covariant derivative $D_{\nu}=\partial_\nu -i e A_{\nu}$ includes the external gauge field
$A_{\nu}$. 

The structure of the (inverse) fermion propagator is given by the following ansatz:
\begin{equation}
iG^{-1}(u,u^\prime) =\Big[(i\partial_t+\mu)\gamma^0 -
(\bm{\pi}_{\perp}\cdot\bm{\gamma})-\pi^{3}\gamma^3
+ i\tilde{\mu}\gamma^1\gamma^2
+\Delta\gamma^3\gamma^5
-m\Big]\delta^{4}(u- u^\prime),
\label{ginverse}
\end{equation}
where $u=(t,\mathbf{r})$ and the canonical momenta are $\pi_{\perp}^{k} \equiv i \partial^k + e A^k$
(with $k=1,2$) and $\pi^{3} = i \partial^3 =- i \partial_3$. While the spatial components of the gradient 
$\bm{\nabla}$ are given by covariant components $\partial_k$, the spatial components of the vector 
potential $\mathbf{A}$ are identified with the contravariant components $A^k$. We choose the vector 
potential in the Landau gauge, $\mathbf{A}= (0, x B,0)$, where $B$ is the strength of the magnetic field
pointing in the $z$-direction.

In Eq.~(\ref{ginverse}), in addition to the usual tree level terms, two new dynamical parameters 
($\tilde{\mu}$ and $\Delta$) are included. From the Dirac structure, it should be clear that 
$\tilde{\mu}$ plays the role of an anomalous magnetic moment and $\Delta$ is the chiral 
shift parameter. In the mean-field approximation, there are no solutions with a nontrivial 
$\tilde{\mu}$ \cite{Gorbar:2011ya}. So, we take $\tilde{\mu}\equiv 0$ below. Note that in $2 + 1$
dimensions (without $z$ coordinate), $\Delta\gamma^3\gamma^5$ would be a mass term 
that is odd under time reversal. This mass is responsible for inducing the Chern-Simons term 
in the effective action for gauge fields \cite{CS}, and it plays an important role in the quantum 
Hall effect in graphene \cite{GGMS2008}.

{\bf Gap equation.} In the mean-field approximation, the gap equation is equivalent to 
the following set of three equations \cite{Gorbar:2011ya,Gorbar:2010}:
\begin{eqnarray}
\mu =\mu_0 -\frac{1}{2}G_{\rm int} \langle j^{0}\rangle  ,
\qquad
m = m_0 - G_{\rm int}  \langle \bar{\psi}\psi\rangle     ,
\qquad
\Delta = -\frac{1}{2}G_{\rm int}  \langle j^{3}_5\rangle   ,
\label{gap-Delta-text} 
\end{eqnarray}
which are solved to determine the three dynamical parameters $\mu$, $m$, and $\Delta$.
Here we will not discuss the vacuum solution, realized at small values of the chemical potential 
($\mu_0\lesssim m_{\rm dyn}/\sqrt{2}$) as the result of the magnetic catalysis \cite{MC1}, but 
concentrate exclusively on the normal ground state with $\Delta\neq 0$, which occurs at nonzero 
fermion density. 

Let us start by analyzing the equation for $\Delta$ in perturbation theory. In the zero order 
approximation, $\mu=\mu_0$ and $\Delta=0$, while the fermion number density $\langle j^0\rangle$ 
and the axial current density $\langle j_5^3\rangle$ are nonzero. In particular, as discussed in 
Ref.~\cite{Metlitski:2005pr}, $\langle j^3_5\rangle_0 =-eB\mu_0/(2\pi^2)$.
(Our convention is such that the electric charge of the electron is $-e$ where $e>0$.) 
To the leading order in the coupling constant, one finds from Eq.~(\ref{gap-Delta-text}) that 
$\Delta \propto G_{\rm int} \langle j^3_5\rangle_0 \neq 0$ and  $\mu - \mu_0 \propto 
G_{\rm int}\langle j^0\rangle_0 \neq 0$. The latter implies that $\mu$ and $\mu_0$ are 
nonequal (in the model at hand, this is a consequence of the Hartree contribution to 
the gap equation). More importantly, we find that a nonzero $\Delta$ is induced. The same 
conclusion is also reached in a more careful analysis of the gap equations, utilizing a 
proper-time regularization \cite{Gorbar:2011ya}. This finding has interesting implications 
for theory and applications. 

{\bf Axial current density.}
As pointed out in Refs.~\cite{Son-et-al,Metlitski:2005pr}, the structure of the topological axial current, 
induced at the LLL, is intimately connected with the axial anomaly \cite{ABJ}. Then, the important 
question is whether the form of the induced axial current $\langle j^{3}_{5}\rangle$  coincides with 
the result in the theory of noninteracting fermions \cite{Son-et-al,Metlitski:2005pr,Rebhan,Hong:2010hi}, 
or whether it is affected by interactions.

We find that the dynamical generation of the chiral shift parameter $\Delta$ does modify the ground 
state expectation value of the axial current density  \cite{Gorbar:2009bm,Gorbar:2011ya,Gorbar:2010}. 
The corresponding correction to the current density was calculated using several different regularization 
schemes (including the gauge invariant proper-time and point-splitting ones \cite{Gorbar:2011ya,
Gorbar:2010}). It reads $\langle j_5^3\rangle -\langle j_5^3\rangle_0 \propto a\Lambda^2 \Delta $, 
where $\Lambda$ is a cut-off parameter and $a$ is a dimensionless constant of order $1$.
Formally, this contribution to the the axial current appears to be quadratically divergent 
when $\Lambda\to \infty$. However, it is finite because the solution for $\Delta$ itself
is inversely proportional to $\Lambda^2$. After taking this into account, one finds that the axial 
current density is finite in the continuum limit. The same is expected in renormalizable gauge 
theories, in which $\Delta$ will be a running parameter that falls off quickly enough in 
ultraviolet to render a finite (or, perhaps, even vanishing) correction to the axial current.

{\bf Axial anomaly relation.}
The above result for the axial current density, which gets a correction due to the chiral shift parameter, 
brings up the question whether the conventional axial anomaly relation \cite{ABJ} is affected in any 
way. This issue was studied in Ref.~\cite{Gorbar:2010}, using a gauge invariant point-splitting 
regularization scheme, and it was found that the chiral shift parameter does not modify the 
axial anomaly. This is in agreement with the findings of Refs.~\cite{Son-et-al,Metlitski:2005pr}.

\section{Applications}

\begin{figure}
\begin{center}
\includegraphics[width=.27\textwidth]{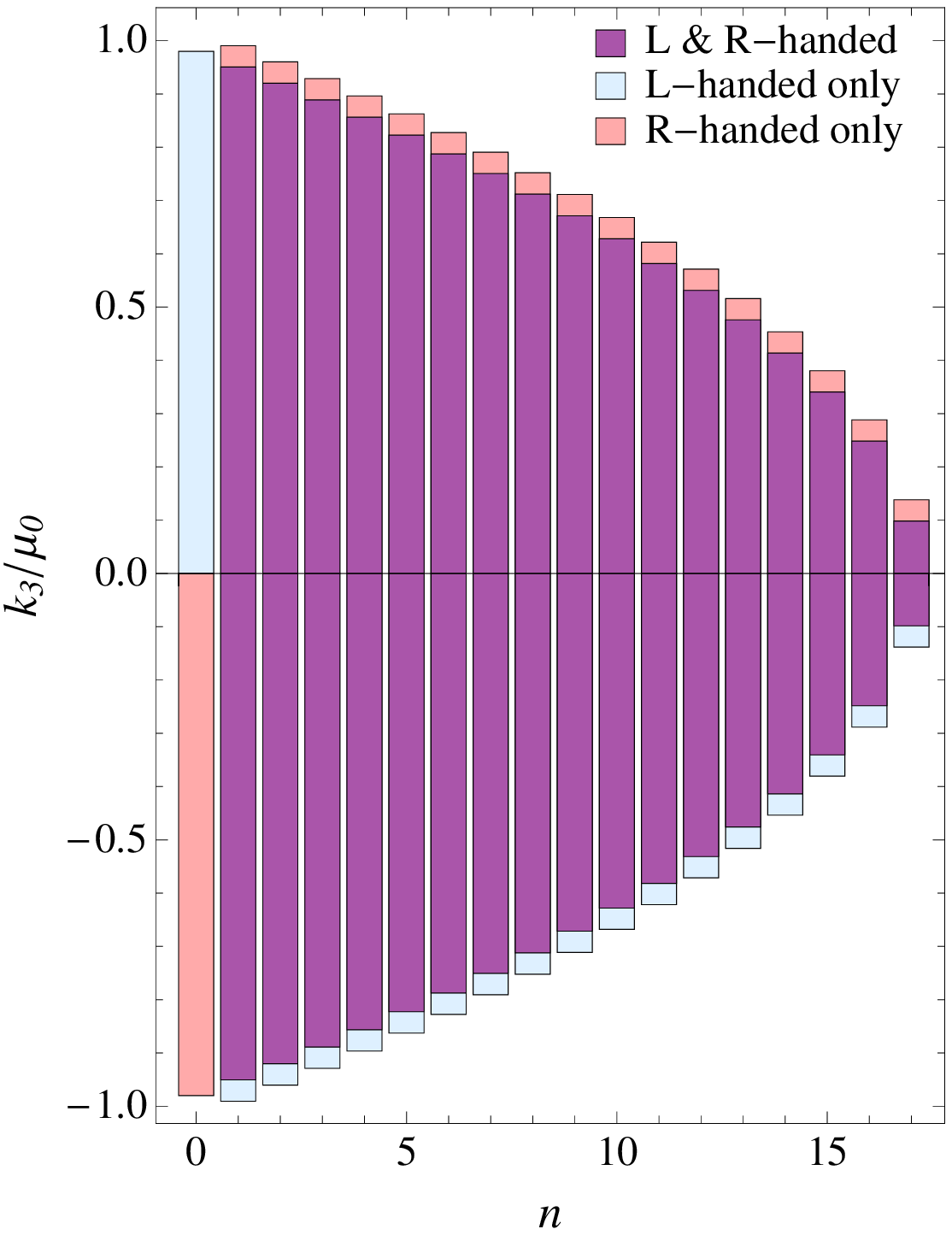}\hspace{0.1\textwidth}
\includegraphics[width=.6\textwidth]{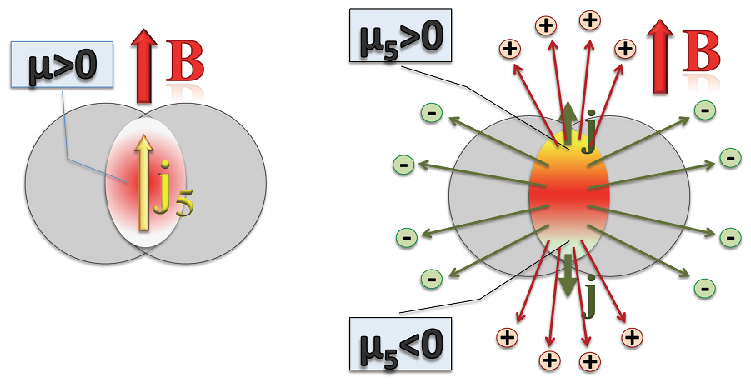}
\caption{Left panel: A schematic distribution of (negatively charged) particles in the ground 
state of cold dense relativistic matter in a magnetic field, which is relevant for physics of 
compact stars.  Middle and right panels: The realization of the modified chiral magnetic 
effect in heavy ion collisions, when an axial current is initially driven by a nonzero baryon 
chemical potential that results in two back-to-back electric currents produced by the axial charges 
in the polar regions.}
\label{figsCombined}
\end{center}
\end{figure}

{\bf Fermi surface.} The immediate implication of a nonzero chiral shift parameter in dense magnetized 
matter is the modification of the quasiparticle dispersion relations. These relations can be used to 
determine the Fermi surface in the space of the longitudinal momentum $k^3$ and the Landau index 
$n$. In relativistic dense matter ($\mu\gg m$), the corresponding states at the Fermi surface can be 
approximately characterized by their chiralities. Taking this into account, it is possible to define 
quasiparticles at the Fermi surface, which are predominantly left-handed or right-handed. Without 
loss of generality, let us assume that $s_{\perp}=\mbox{sgn}(eB)>0$. Then, the Fermi surface for the 
{\em predominantly left-handed} particles is given by 
\begin{eqnarray}
n=0: &\!\!\!\!& k^{3}=+\sqrt{(\mu-s_\perp \Delta)^2-m^2},   \label{Fermi-k3-1L}\\
n>0: &\!\!\!\!& k^{3}=+\sqrt{\left(\sqrt{\mu^2-2n|eB|}- s_\perp \Delta\right)^2-m^2}, \label{Fermi-k3-2L} \\
        &\!\!\!\!& k^{3}=-\sqrt{\left(\sqrt{\mu^2-2n|eB|}+ s_\perp \Delta\right)^2-m^2}, \label{Fermi-k3-3L}
\end{eqnarray}
and the Fermi surface for the {\em predominantly right-handed} particles is 
\begin{eqnarray}
n=0: &\!\!\!\!& k^{3}=-\sqrt{(\mu-s_\perp \Delta)^2-m^2},   \label{Fermi-k3-1R}\\
n>0: &\!\!\!\!& k^{3}=-\sqrt{\left(\sqrt{\mu^2-2n|eB|}- s_\perp \Delta\right)^2-m^2}, \label{Fermi-k3-2R}\\
        &\!\!\!\!& k^{3}=+\sqrt{\left(\sqrt{\mu^2-2n|eB|}+ s_\perp \Delta\right)^2-m^2}. \label{Fermi-k3-3R}
\end{eqnarray}
In the massless case, of course, this correspondence becomes exact. Then, we find that the Fermi 
surface for fermions of a given chirality is asymmetric in the direction of the magnetic field. In the 
left panel in Fig.~\ref{figsCombined}, we show a schematic distribution of negatively charged 
fermions and take into account that the parameter $s_\perp \Delta$ has the same sign as the chemical potential. 
(A similar distribution is also valid for positively charged fermions, but the left-handed and right-handed fermions 
will interchange their roles.) For the fermions of a given chirality, the LLL and the higher Landau levels give 
opposite contributions to the overall asymmetry of the Fermi surface. For example, the left-handed electrons in 
the LLL occupy only the states with {\em positive} longitudinal momenta (pointing in the magnetic field direction). The 
spins of the corresponding LLL electrons point against the magnetic field direction. In the higher Landau levels, 
while the left-handed electrons can have both positive and negative longitudinal momenta (as well as both spin 
projections), there are more states with {\em negative} momenta occupied, see Fig.~\ref{figsCombined}. 
If there are many Landau levels occupied, which is the case when $\mu\gg \sqrt{|eB|}$, the relative contribution 
of the LLL to the whole Fermi surface is small, and the overall asymmetry is dominated by higher Landau levels.
In the opposite regime of super-strong magnetic field (if it can be realized in compact stars at all), only the LLL is 
occupied and, therefore, the overall asymmetry of the Fermi surface will be reversed. In the intermediate regime 
of a few Landau levels occupied, one should expect a crossover from one regime to the other, where the 
asymmetry goes through zero.

{\bf Compact stars.}
The asymmetry with respect to longitudinal momentum $k^{3}$ of the opposite chirality fermions 
in the ground state of dense magnetized matter, discussed above, may have important physical 
consequences. For example, the fact that only the left-handed fermions participate in the weak 
interactions means that the neutrinos will scatter asymmetrically off the matter, in which the chiral 
shift parameter is nonvanishing. 

By making use of this observation, a qualitatively new mechanism for the pulsar kicks \cite{pulsarkicks} 
was proposed in Ref.~\cite{Gorbar:2009bm}. It can be realized in almost any type of relativistic matter 
inside a protoneutron star (e.g., the electron plasma of the nuclear/hadronic matter, or the quark and 
electron plasma in the deconfined quark matter), in which a nonzero chiral shift parameter $\Delta$ 
develops. 

When trapped neutrinos gradually diffuse through the bulk of a protoneutron star, they can {\em build up} 
an asymmetric momentum distribution as a result of their multiple elastic scattering on the {\em nonisotropic} 
state of left-handed fermions (electrons or quarks). This is in contrast to the common dynamics of diffusion 
through an {\em isotropic} hot matter, which leads to a very efficient thermal isotropization and, therefore, 
a wash out of any original nonisotropic distribution of neutrinos \cite{Kusenko,SagertSchaffner}.

It appears also very helpful for the 
new pulsar kick mechanism that the chiral shift parameter is not much affected even by moderately 
high temperatures, $10~\mbox{MeV} \lesssim T\lesssim 50~\mbox{MeV}$, present during the earliest
stages of protoneutron stars \cite{AnnRevNuclPart}. Indeed, as our findings show, the value of 
$\Delta$ is primarily determined by the chemical potential and has a weak/nonessential temperature 
dependence when $\mu\gg T$. In the stellar context, this ensures the feasibility of the proposed mechanism 
even at the earliest stages of the protoneutron stars, when there is sufficient amount of thermal energy 
to power the strongest (with $v\gtrsim 1000~\mbox{km/s}$) pulsar kicks observed \cite{pulsarkicks}.
Alternatively, the constraints of the energy conservation would make it hard, if not impossible, to explain 
any sizable pulsar kicks if the interior matter is cold ($T\lesssim 1~\mbox{MeV}$).
 
Let us also mention that the robustness of the chiral shift in hot magnetized matter may be useful to 
provide an additional neutrino push to facilitate successful supernova explosions as suggested in 
Ref.~\cite{Fryer:2005sz}. The specific details of such a scenario are yet to be worked out.

%%%%%%%%%%%%%%%%%%%%%%%%%%%%%%%%%%%%%%%
%%%%%%%%%%%%%%%%%%%%%%%%%%%%%%%%%%%%%%%
%%%%%%%%%%%%%%%%%%%%%%%%%%%%%%%%%%%%%%%
%%%%%%%%%%%%%%%%%%%%%%%%%%%%%%%%%%%%%%%

{\bf Heavy Ion Collisions.} It is natural to ask whether the chiral shift parameter can have any interesting 
implications in the regime of relativistic heavy ion collisions, where sufficiently strong magnetic fields may 
exist \cite{Skokov:2009qp}.  The examples of the recently suggested phenomena, that appear to be 
closely related to the generation of the chiral shift, are the chiral magnetic effect \cite{Kharzeev:2007tn,Rebhan}, 
the chiral magnetic spiral \cite{Basar:2010zd,Kim,Frolov}, and the chiral magnetic wave \cite{Kharzeev}.

As was recently revealed in Ref. \cite{Gorbar:2011ya}, at high temperatures, i.e., in the regime relevant for relativistic 
heavy ion collisions, the chiral shift parameter is not suppressed and generated for any nonzero chemical potential. 
However, its role is not as obvious as in the case of stellar matter. At high temperatures, the Fermi surface and the 
low-energy excitations in its vicinity are not very useful concepts any more. Instead, it is the axial current itself that 
is of interest. The chiral shift should induce a correction to the topological axial current, see the middle panel in 
Fig.~\ref{figsCombined}. However, unlike the topological term, which is also proportional to the chemical potential, 
the dynamical one contains an extra factor of the coupling constant. Therefore, only at relatively strong coupling, 
which can be provided by QCD interactions, the effect of the chiral shift parameter on the axial current can be substantial.

Following the ideas similar to those that were used in the chiral magnetic effect \cite{Kharzeev:2007tn,Rebhan,
Fukushima:2010zza}, we have recently suggested \cite{Gorbar:2011ya} that the axial current by itself can 
play an important role in hot matter produced by heavy ion collisions and lead to a modified version of the 
chiral magnetic effect, which does not rely on the initial topological charge fluctuations \cite{Kharzeev:2007tn}. 
An initial axial current generates an excess of opposite chiral charges around the polar regions of the fireball. 
Then, these chiral charges trigger two ``usual" chiral magnetic effects with opposite directions of the vector 
currents at the opposite poles (see the right panel in Fig. \ref{figsCombined}). 
The inward flows of these electric currents will diffuse inside the fireball, 
while the outward flows will lead to a distinct observational signal: an excess of same sign charges going 
back-to-back. A numerical estimate of the modified chiral magnetic effect has been recently done in 
Ref.~\cite{Burnier:2011bf}.

Concerning the regime of hot relativistic matter, let us also mention that it will be of interest to extend 
our analysis of magnetized relativistic matter to address the properties of collective modes, similar to 
those presented in Ref.~\cite{Kharzeev}, by studying various current-current correlators.

\section{Outlook}

The present analysis was performed in the framework of the NJL model. It would be 
important to extend it to realistic field theories, QED and QCD. In connection with
that, we would like to note the following. The expression for the chiral shift parameter,
$\Delta \sim g\mu \, eB/\Lambda^2$, obtained in the NJL model implies that both fermion
density and magnetic field are necessary for the generation of $\Delta$. This feature
should also be valid in renormalizable theories. As for the cutoff $\Lambda$, it enters
the results only because of the nonrenormalizability of the NJL model.

Similar studies of chiral
symmetry breaking in the vacuum ($\mu_0 = 0$) QED and QCD in a magnetic field show
that the cutoff scale $\Lambda$ is replaced by $\sqrt{|eB|}$ there \cite{QED}. Therefore,
one might expect that in QED and QCD with both $\mu$ and $B$ being nonzero, $\Lambda$
will be replaced by a physical parameter, such as $\sqrt{|eB|}$. This in turn suggests that
a constant chiral shift parameter $\Delta$ will become a running quantity that depends on the
longitudinal momentum $k^3$ and the Landau level index $n$.

Another important feature that one could expect in QCD in a magnetic field is a topological
contribution in the baryon charge \cite{SS} connected with collective massless fermion
excitations in the phase with spontaneous chiral symmetry breaking. This feature could
dramatically change the properties of that phase \cite{Rebhan}.

It is clear that dynamics in dense relativistic matter in a magnetic field is rich and
sophisticated. In particular, one could expect surprises in studies of the phase diagram
of QCD in a magnetic field \cite{Gatto:2010qs,Mizner,Fayazbakhsh:2010bh,D'Elia}.

\section*{Acknowledgments}
V.A.M. is grateful to the organizers of the  International school of nuclear physics, 33rd course, 
Erice-Sicily, in particular, Prof. Armand Faessler and Prof. Jochen Wambach, for their warm 
hospitality. The work of E.V.G. was supported partially by the SCOPES under 
Grant No. IZ73Z0-128026 of the Swiss NSF, under Grant No. SIMTECH 246937 
of the European FP7 program, the joint Grant RFFR-DFFD No. F28.2/083 of the 
Russian Foundation for Fundamental Research and of the Ukrainian State Foundation 
for Fundamental Research (DFFD). The work of V.A.M. was supported by the Natural 
Sciences and Engineering Research Council of Canada. The work of I.A.S. is supported 
by the U.S. National Science Foundation under Grant No. PHY-0969844.


\begin{thebibliography}{99}
\itemsep -2pt 

\bibitem{Son-et-al}
    D.~T.~Son and A.~R.~Zhitnitsky,
  %``Quantum anomalies in dense matter,''
  %%CITATION = PHRVA,D70,074018;%%
  \Journal{\PRD}{70}{074018}{2004};
  %[hep-ph/0405216].
    G.~M.~Newman and D.~T.~Son,
  %``Response of strongly-interacting matter to magnetic field: Some exact results,''
  \Journal{\PRD}{73}{045006}{2006}.
 % [hep-ph/0510049].
   %%CITATION = PHRVA,D73,045006;%%

\bibitem{Metlitski:2005pr}
  M.~A.~Metlitski and A.~R.~Zhitnitsky,
  %``Anomalous axion interactions and topological currents in dense matter,''
  \Journal{\PRD}{72}{045011}{2005}.
  %%CITATION = PHRVA,D72,045011;%%
  
\bibitem{Kharzeev:2007tn}
  D.~Kharzeev,
  %``Parity violation in hot QCD: Why it can happen, and how to look for it,''
  \Journal{\PLB}{633}{260}{2006};
  %[arXiv:hep-ph/0406125 [hep-ph]].
  %%CITATION = PHLTA,B633,260;%%
    D.~Kharzeev and A.~Zhitnitsky,
  %``Charge separation induced by P-odd bubbles in QCD matter,''
  \Journal{\NPA}{797}{67}{2007};
%  [arXiv:0706.1026 [hep-ph]].
  %%CITATION = NUPHA,A797,67;%%
  D.~E.~Kharzeev, L.~D.~McLerran and H.~J.~Warringa,
  %``The Effects of topological charge change in heavy ion collisions: 'Event by event P and CP violation',''
  \Journal{\NPA}{803}{227}{2008};
% [arXiv:0711.0950 [hep-ph]].
  %%CITATION = NUPHA,A803,227;%%
  K.~Fukushima, D.~E.~Kharzeev and H.~J.~Warringa,
  %``The Chiral Magnetic Effect,''
  \Journal{\PRD}{78}{074033}{2008};
  %%CITATION = PHRVA,D78,074033;%%
  Seung-il~Nam,
  %``Vector current correlation and charge separation via chiral magnetic
  %effect,''
  \Journal{\PRD}{82}{045017}{2010}.
  %[arXiv:1004.3444 [hep-ph]].
  %%CITATION = PHRVA,D82,045017;%%
  
\bibitem{Gorbar:2009bm}
  E.~V.~Gorbar, V.~A.~Miransky, and I.~A.~Shovkovy,
  %``Chiral asymmetry of the Fermi surface in dense relativistic matter in a
  %magnetic field,''
  \Journal{\PRC}{80}{032801(R)}{2009}.
  %%CITATION = ARXIV:0904.2164;%%

\bibitem{Rebhan}
  A.~Rebhan, A.~Schmitt and S.~A.~Stricker,
  %``Anomalies and the chiral magnetic effect in the Sakai-Sugimoto model,''
  \Journal{\JHEP}{01}{026}{2010};
  %%CITATION = JHEPA,1001,026;%%
  F.~Preis, A.~Rebhan and A.~Schmitt,
  %``Inverse magnetic catalysis in dense holographic matter,''
  \Journal{\JHEP}{03}{033}{2011}.
  %arXiv:1012.4785 [hep-th].
  %%CITATION = ARXIV:1012.4785;%%

\bibitem{FI1}
  D.~Ebert, K.~G.~Klimenko, M.~A.~Vdovichenko, and A.~S.~Vshivtsev,
  %``Magnetic oscillations in dense cold quark matter with four-fermion
  %interactions,''
  \Journal{\PRD}{61}{025005}{1999};
  %% [arXiv:hep-ph/9905253].
  %%CITATION = PHRVA,D61,025005;%%
  E.~J.~Ferrer, V.~de la Incera, and C.~Manuel,
  %``Magnetic color flavor locking phase in high density QCD,''
  \Journal{\PRL}{95}{152002}{2005};
  %%CITATION = PRLTA,95,152002;%%
  E.~J.~Ferrer and V.~de la Incera,
  %``Magnetic phases in three-flavor color superconductivity,''
  \Journal{\PRD}{76}{045011}{2007};
  %%CITATION = PHRVA,D76,045011;%%
  K.~Fukushima and H.~J.~Warringa,
 %``Color superconducting matter in a magnetic field,''
  \Journal{\PRL}{100}{032007}{2008};
  %%CITATION = PRLTA,100,032007;%%
  J.~L.~Noronha and I.~A.~Shovkovy,
  %``Color-flavor locked superconductor in a magnetic field,''
  \Journal{\PRD}{76}{105030}{2007}.
  %%CITATION = PHRVA,D76,105030;%%

\bibitem{SS} D.~T.~Son and M.~A.~Stephanov,
  %``Axial anomaly and magnetism of nuclear and quark matter,''
  \Journal{\PRD}{77}{014021}{2008}.
  %%CITATION = PHRVA,D77,014021;%%

\bibitem{Basar:2010zd}
  G.~Basar, G.~V.~Dunne and D.~E.~Kharzeev,
  %``Chiral Magnetic Spirals,''
  \Journal{\PRL}{104}{232301}{2010}.
  %%CITATION = PRLTA,104,232301;%%

\bibitem{Kim}
  K.~Y.~Kim, B.~Sahoo and H.~U.~Yee,
  %``Holographic chiral magnetic spiral,''
  \Journal{\JHEP}{10}{005}{2010}.
  %%CITATION = JHEPA,1010,005;%%
%\bibitem{Kharzeev:2010gd}

\bibitem{Frolov}
  I.~E.~Frolov, V.~C.~Zhukovsky and K.~G.~Klimenko,
  %``Chiral density waves in quark matter within the Nambu--Jona-Lasinio model
  %in an external magnetic field,''
  \Journal{\PRD}{82}{076002}{2010}.
  %%CITATION = PHRVA,D82,076002;%%

\bibitem{Osipov:2007je}
  A.~A.~Osipov, B.~Hiller, A.~H.~Blin, J.~da Provid\^{e}ncia,
  %``Dynamical chiral symmetry breaking by a magnetic field and multi-quark interactions,''
  \Journal{\PLB}{650}{262}{2007}.
  %[hep-ph/0701090].
  %%CITATION = PHLTA,B650,262;%%

\bibitem{Gatto:2010qs}
  R.~Gatto and M.~Ruggieri,
  %``Dressed Polyakov loop and phase diagram of hot quark matter under magnetic
  %field,''
  \Journal{\PRD}{82}{054027}{2010}.
  %[arXiv:1007.0790 [hep-ph]].
  %%CITATION = PHRVA,D82,054027;%%
%\bibitem{Mizher}

\bibitem{Mizner} 
  A.~J.~Mizher, M.~N.~Chernodub and E.~S.~Fraga,
  %``Phase diagram of hot QCD in an external magnetic field: possible splitting
  %of deconfinement and chiral transitions,''
  \Journal{\PRD}{82}{105016}{2010}.
  %%CITATION = PHRVA,D82,105016;%%

\bibitem{Fayazbakhsh:2010bh}
  S.~Fayazbakhsh and N.~Sadooghi,
  %``Phase diagram of hot magnetized two-flavor color superconducting quark matter,''
  \Journal{\PRD}{83}{025026}{2011}.
  % [arXiv:1009.6125 [hep-ph]].
   %%CITATION = PHRVA,D83,025026;%%

\bibitem{Buividovich:2009wi}
P.~V.~Buividovich, M.~N.~Chernodub, E.~V.~Luschevskaya and M.~I.~Polikarpov,
%``Numerical evidence of chiral magnetic effect in lattice gauge theory,''
\Journal{\PRD}{80}{054503}{2009};
%[arXiv:0907.0494 [hep-lat]].
%%CITATION = PHRVA,D80,054503;%%
%\bibitem{Buividovich:2008wf}
% P.~V.~Buividovich, M.~N.~Chernodub, E.~V.~Luschevskaya and M.~I.~Polikarpov,
%``Numerical study of chiral symmetry breaking in non-Abelian gauge theory with background magnetic field,''
\Journal{\PLB}{682}{484}{2010};
%[arXiv:0812.1740 [hep-lat]].
%%CITATION = ARXIV:0812.1740;%%
%\bibitem{Buividovich:2010tn}
  P.~V.~Buividovich, M.~N.~Chernodub, D.~E.~Kharzeev, T.~Kalaydzhyan, E.~V.~Luschevskaya, M.~I.~Polikarpov,
  %``Magnetic-Field-Induced insulator-conductor transition in SU(2) quenched lattice gauge theory,''
  \Journal{\PRL}{105}{132001}{2010}.
  %[arXiv:1003.2180 [hep-lat]].
  %%CITATION = PRLTA,107,132001;%%

\bibitem{D'Elia}
  M.~D'Elia, S.~Mukherjee and F.~Sanfilippo,
  %``QCD Phase Transition in a Strong Magnetic Background,''
  \Journal{\PRD}{82}{051501}{2010};
  %%CITATION = PHRVA,D82,051501;%%
    M.~D'Elia and F.~Negro,
  %``Chiral Properties of Strong Interactions in a Magnetic Background,''
  \Journal{\PRD}{83}{114028}{2011};
%  [arXiv:1103.2080 [hep-lat]].
  %%CITATION = PHRVA,D83,114028;%%
%  \bibitem{Bruckmann:2011zx}
  F.~Bruckmann and G.~Endrodi,
  %``Dressed Wilson loops as dual condensates in response to magnetic and
  %electric fields,''
  \Journal{\PRD}{84}{074506}{2011}.
  %[arXiv:1104.5664 [hep-lat]].
  %%CITATION = PHRVA,D84,074506;%%
  
  \bibitem{arXiv:1111.4956} 
  G.~S.~Bali, F.~Bruckmann, G.~Endrodi, Z.~Fodor, S.~D.~Katz, S.~Krieg, A.~Schafer and K.~K.~Szabo,
  %``The QCD phase diagram for external magnetic fields,''
  arXiv:1111.4956 [hep-lat].
  %%CITATION = ARXIV:1111.4956;%%
  
\bibitem{Gorbar:2011ya}
  E.~V.~Gorbar, V.~A.~Miransky, I.~A.~Shovkovy,
  %``Normal ground state of dense relativistic matter in a magnetic field,''
  \Journal{\PRD}{83}{085003}{2011}.
%  [arXiv:1101.4954 [hep-ph]].
  %%CITATION = PHRVA,D83,085003;%%

\bibitem{CS}
  A.~J.~Niemi and G.~W.~Semenoff,
  %``Axial Anomaly Induced Fermion Fractionization And Effective Gauge Theory
  %Actions In Odd Dimensional Space-Times,''
  \Journal{\PRL}{51}{2077}{1983};
  %%CITATION = PRLTA,51,2077;%%
  A.~N.~Redlich,
  %``GAUGE NONINVARIANCE AND PARITY NONCONSERVATION OF THREE-DIMENSIONAL
  %FERMIONS,''
  \Journal{\PRL}{52}{18}{1984};
  %%CITATION = PRLTA,52,18;%%
 %``Parity Violation And Gauge Noninvariance Of The Effective Gauge Field
  %Action In Three-Dimensions,''
  \Journal{\PRD}{29}{2366}{1984}.
  %%CITATION = PHRVA,D29,2366;%%

\bibitem{GGMS2008}
E.~V.~Gorbar, V.~P.~Gusynin, V.~A.~Miransky, and I.~A.~Shovkovy,
  %``Dynamics in the quantum Hall effect and the phase diagram of graphene,''
  \Journal{\PRB}{78}{085437}{2008};
  %%CITATION = PHRVA,B78,085437;%%
E.~V.~Gorbar, V.P.~Gusynin, and V.~A.~Miransky,
  \Journal{\LTP}{34}{790}{2008}.
%%CITATION = LTPHE,34,790;%%

\bibitem{Gorbar:2010}
  E.~V.~Gorbar, V.~A.~Miransky, and I.~A.~Shovkovy,
  %``Chiral asymmetry and axial anomaly in magnetized relativistic matter,''
  \Journal{\PLB}{695}{354}{2011}.
  %%CITATION = PHLTA,B695,354;%%

\bibitem{MC1} V. P.~Gusynin, V. A.~Miransky, and I. A.~Shovkovy,
\Journal{\PRL}{73}{3499}{1994};
  %%CITATION = PRLTA,73,3499;%%
\Journal{\PLB}{349}{477}{1995}.
%%CITATION = PHLTA,B349,477;%%

\bibitem{ABJ}
  S.~L.~Adler,
  %``Axial vector vertex in spinor electrodynamics,''
  \Journal{\PREV}{177}{2426}{1969};
  %%CITATION = PHRVA,177,2426;%%
    J.~S.~Bell and R.~Jackiw,
  %``A PCAC puzzle: pi0 $\to$ gamma gamma in the sigma model,''
  \Journal{\NCA}{60}{47}{1969}.
  %%CITATION = NUCIA,A60,47;

\bibitem{Hong:2010hi}
  V.~A.~Rubakov,
  %``On chiral magnetic effect and holography,''
  arXiv:1005.1888 [hep-ph];
  %%CITATION = ARXIV:1005.1888;%%
  D.~K.~Hong,
  %``Anomalous currents in dense matter under a magnetic field,''
  \Journal{\PLB}{699}{305}{2011}.
%  [arXiv:1010.3923 [hep-th]].
%%CITATION = PHLTA,B699,305;%%

\bibitem{pulsarkicks}
  A.~G.~Lyne and D.~R.~Lorimer,
  %``High birth velocities of radio pulsars,''
  \Journal{\Nature}{369}{127}{1994};
  %%CITATION = NATUA,369,127;%%
  J.~M.~Cordes and D.~F.~Chernoff,
  %``Neutron Star Population Dynamics.II: 3D Space Velocities of Young
  %Pulsars,''
  \Journal{\APJ}{505}{315}{1998};
  %% [arXiv:astro-ph/9707308]
  %%CITATION = ASJOA,505,315;%%
  B.~M.~S.~Hansen and E.~S.~Phinney,
  %``The Pulsar Kick Velocity Distribution,''
  \Journal{\MNRAS}{291}{569}{1997};
  %% [arXiv:astro-ph/9708071].
  %%CITATION = MNRAA,291,569;%%
  C.~Fryer, A.~Burrows, and W.~Benz,
  %``Population Synthesis for Neutron Star Systems with Intrinsic Kicks,''
  \Journal{\APJ}{496}{333}{1998};
  %% [arXiv:astro-ph/9710333].
  %%CITATION = ASJOA,496,333;%%
  Z.~Arzoumanian, D.~F.~Chernoffs, and J.~M.~Cordes,
  %``The Velocity Distribution of Isolated Radio Pulsars s,''
  \Journal{\APJ}{568}{289}{2002}.
  %% [arXiv:astro-ph/0106159].
  %%CITATION = ASJOA,568,289;%%

\bibitem{Kusenko} A.~Kusenko, G.~Segre, and A.~Vilenkin,
  %``Neutrino transport: No asymmetry in equilibrium,''
  \Journal{\PLB}{437}{359}{1998}.
  %% [arXiv:astro-ph/9806205].
  %%CITATION = PHLTA,B437,359;%%

\bibitem{SagertSchaffner}
  I.~Sagert and J.~Schaffner-Bielich,
  %``Pulsar kicks by anisotropic neutrino emission from quark matter,''
   \Journal{\JPG}{35}{014062}{2008};
  %%CITATION = JPHGB,G35,014062;%%
%  I.~Sagert and J.~Schaffner-Bielich,
  %``Pulsar kicks by anisotropic neutrino emission from quark matter in strong
  %magnetic fields,''
\Journal{\AA}{489}{281}{2008}.
  %arXiv:0708.2352 [astro-ph].
  %%CITATION = ARXIV:0708.2352;%%

\bibitem{AnnRevNuclPart}
D.~Page and S.~Reddy,
  %``Dense Matter in Compact Stars: Theoretical Developments and Observational
  %Constraints,''
  \Journal{\ARNPS}{56}{327}{2006}.
  %%[arXiv:astro-ph/0608360].
  %%CITATION = ARNUA,56,327;%%

\bibitem{Fryer:2005sz}
  C.~L.~Fryer and A.~Kusenko,
  %``Effects of neutrino-driven kicks on the supernova explosion mechanism,''
  \Journal{\APJS}{163}{335}{2006}.
  %% [arXiv:astro-ph/0512033].
  %%CITATION = APJSA,163,335;%%

\bibitem{Skokov:2009qp}
  V.~Skokov, A.~Illarionov and V.~Toneev,
  %``Estimate of the magnetic field strength in heavy-ion collisions,''
  \Journal{\IJMPA}{24}{5925}{2009}.
  %arXiv:0907.1396 [nucl-th].
  %%CITATION = ARXIV:0907.1396;%%

\bibitem{Kharzeev}
  D.~E.~Kharzeev and H.~U.~Yee,
  %``Chiral Magnetic Wave,''
  \Journal{\PRD}{83}{085007}{2011}.
 % [arXiv:1012.6026 [hep-th]].
  %%CITATION = PHRVA,D83,085007;%%

\bibitem{Fukushima:2010zza}
  K.~Fukushima and M.~Ruggieri,
  %``Dielectric correction to the Chiral Magnetic Effect,''
   \Journal{\PRD}{82}{054001}{2010}.
  %[arXiv:1004.2769 [hep-ph]].
  %%CITATION = PHRVA,D82,054001;%%
  
\bibitem{Burnier:2011bf}
  Y.~Burnier, D.~E.~Kharzeev, J.~Liao, H.~-U.~Yee,
  %``Chiral magnetic wave at finite baryon density and the electric quadrupole moment of quark-gluon plasma in heavy ion collisions,''
  \Journal{\PRL}{107}{052303}{2011}.
   %[arXiv:1103.1307 [hep-ph]].
  %%CITATION = PRLTA,107,052303;%%

\bibitem{QED}
V.~P.~Gusynin, V.~A.~Miransky and I.~A.~Shovkovy,
  %``Dynamical chiral symmetry breaking by a magnetic field in QED,''
  \Journal{\PRD}{52}{4747}{1995};
  %[arXiv:hep-ph/9501304].
  %%CITATION = PHRVA,D52,4747;%%
%``Theory of the magnetic catalysis of chiral symmetry breaking in QED,''
  \Journal{\NPB}{563}{361}{1999};
  %[arXiv:hep-ph/9908320].
  %%CITATION = NUPHA,B563,361;%%
V.~A.~Miransky and I.~A.~Shovkovy,
  %``Magnetic catalysis and anisotropic confinement in QCD,''
    \Journal{\PRD}{66}{045006}{2002}.
  %[arXiv:hep-ph/0205348].
  %%CITATION = PHRVA,D66,045006;%%

\end{thebibliography}
\end{document}